# Molecular Mechanisms of Polymer Crosslinking via Thermal Activation

Javed Akhtar[a], Jogeswar Chhatria[b], Sooraj Kunnikuruvan[b], Satyesh K. Yadav[c], and Tarak Patra[a]

[a]Department of Chemical Engineering, IIT Madras, Chennai, India – 600036

[b]Department of Chemistry, IIT Madras, Chennai, India - 600036

[c]Department of Metallurgical and Materials Engineering, IIT Madras, Chennai, India – 600036

## Abstract

Developing efficient and universal polymer crosslinking strategies is pivotal for advanced material design, especially for challenging matrixes like polyethylene (PE), polypropylene (PP), and polystyrene (PS). Traditional crosslinkers such as divinylbenzene (DVB) often requires high-temperature radical initiators and are limited by poor compatibility with saturated hydrocarbon matrices. In contrast, bis-diazirine (BD) crosslinkers offer a promising alternative by harnessing thermally or photochemically generated carbene intermediates for highly selective C–H bond insertions. Here, we employ density functional theory (DFT)-based electronic structure calculations to elucidate the molecular mechanisms and energetics of BD-mediated crosslinking across PE, PP, and PS. We demonstrate that BD enables efficient covalent linkage through low free energy barriers (~28–36 kcal/mol), facilitating crosslinking at moderate temperatures without catalysts and with minimal sensitivity to polymer chain length. Moreover, BD exhibits selective reactivity towards the tertiary and secondary C–H bonds in PP and PS, respectively. Comparative analysis shows that BD dramatically outperforms DVB, especially in saturated polymers, enabling reaction times that are orders of magnitude faster. Our findings provide atomistic insights into BD crosslinker reactivity and establish a mechanistic foundation for next-generation, universal C–H activation-based crosslinking technologies.

## I. Introduction

Crosslinked polymers play a vital role in advanced materials design due to their superior mechanical strength, thermal stability, and resistance to solvents and chemical degradation.[1–4] These properties arise from covalent bonding between polymer chains, which forms an extended three-dimensional (3D) network that imparts structural integrity and resilience under mechanical or thermal stress.[5,6] Applications of crosslinked polymers span diverse sectors, including biomedical devices, aerospace composites, coatings, automotive components, and electrical insulation.[7–13] For example, crosslinked polyethylene exhibits superior tensile strength and thermal resistance compared to its linear counterpart, enabling its widespread use in high-voltage cable insulation.[14] Traditional crosslinking approaches often require harsh processing conditions—such as elevated temperatures, high pressures, and/or the use of strong oxidizing agents or radiation sources—to initiate radical or ionic reactions that form covalent interchain bonds.[15–18] For instance, the vulcanization of elastomers—such as natural rubber—requires sulfur and accelerators at temperatures between 720–740 K. This process facilitates the formation of sulfur crosslinks between polymer chains, thereby enhancing elasticity, strength, and thermal resistance of the natural rubber.[19] Similarly, peroxide-initiated crosslinking of polyethylene employs thermal decomposition of peroxides (e.g., dicumyl peroxide) to generate free radicals capable of abstracting hydrogen atoms and initiating C–C bond formation.[20] These processes demand high energy input and are associated with secondary degradation phenomena such as chain scission, branching, and oxidative aging.[21,22] Electron beam (EB) and gamma (γ) irradiation are also industrially employed to induce crosslinking in commodity thermoplastics by cleaving strong C–H bonds (~90–95 kcal/mol), often using energies up to 10 MeV. However, these high-energy methods necessitate specialized equipment and shielding, and they may trigger unwanted side reactions that limit material consistency.[17,23,24] Moreover, not all polymers respond equally to such treatments. Polypropylene (PP), a nonpolar polyolefin with minimal functional groups and over 70 million metric tons produced annually, is especially difficult to crosslink using conventional methods due to its susceptibility to chain degradation under radical-forming conditions.[25,26] Therefore, the difficulty and energy intensity of polymer crosslinking depend on the type of polymer, the crosslinking method, and the reaction conditions. Some polymers have reactive functional groups that naturally undergo crosslinking (e.g., epoxy resins), while others require additional modifications.[27–31] Factors like temperature, pressure, and reaction time must be optimized for efficient crosslinkings.[32,33]

To overcome the above challenges, researchers have explored alternative crosslinking strategies that operate under milder and more controlled conditions. Among them, bis-diazirine (BD) has emerged as a promising candidate, offering unique advantages in terms of reaction selectivity, versatility, and functional group tolerance.[34–38] Lepage et al. have shown BD molecules capable of generating reactive carbenes upon activation with modest thermal (383 – 413 K) or photochemical (UV, ~350 nm) stimuli.[39] These carbenes undergo highly efficient double C–H insertion reactions, facilitating covalent interchain linkages in a wide range of polymer backbones—including challenging substrates like polypropylene (PP), polyethylene (PE), and polystyrene (PS)—without requiring pre-functionalization or external catalysts. This universal approach to polymer crosslinking appears to be a paradigm shift in the design of crosslinking reagents, emphasizing reactivity with ubiquitous C–H bonds over tailored chemical moieties. In contrast, traditional difunctional crosslinkers such as divinylbenzene (DVB) rely on radical polymerization of vinyl groups to form crosslinked networks.[40–42] While effective in systems containing unsaturated or aromatic functionalities (e.g., styrene, acrylates), DVB-based methods typically require high-temperature radical initiators and are not suitable for polyolefins, which lack readily accessible reactive sites.[43–45] Furthermore, the use of DVB can introduce inhomogeneity in network structure due to its relatively slow kinetics and poor compatibility with saturated hydrocarbon matrices.[46–48]

Motivated by the experimental success of BD-based polymer crosslinking, our objective in this study is to elucidate the reaction pathway of polymer crosslinking mediated by BD molecules using density functional theory (DFT)-based electronic structure calculations. By mapping the key intermediates into transition states and quantifying activation barriers, we aim to gain a fundamental understanding of the underlying reaction mechanism and the factors governing its efficiency. We optimize the molecular geometries at two different DFT levels of theory and compute electronic energies at 0 K. Subsequently, we perform a vibrational frequency calculation on the optimized geometries at a target temperature. This provides zero-point energy and thermal correction to energy. We compute the entropy from the vibrational, rotational and translation modes of the molecules. These data are utilized to construct free energy profiles of crosslink reactions. Our analysis clearly suggests that the BD can crosslink polymers in moderate temperatures and without requiring a catalyst. We show a significant energy difference between secondary and tertiary carbon crosslinking of polymers. Furthermore, we examine how the crosslinking energy barrier varies with the polymer chain length. The work provides a mechanistic understanding of polymer crosslinks using BD crosslinkers, and demonstrate their superiority over other traditional crosslinks, such as DVB.

## II. Model and Methodology

Model structures of the polymers of varying repeat units are built using GaussView 6.0.16.[49] We choose three commodity polymers viz., polyethylene (PE), polypropylene (PP), and polystyrene (PS) for this study. DFT calculations are carried out using the Gaussian16 software package, and GaussView 6.0.16 is used for visualization.[50] Geometry optimizations are carried out using both the B3LYP and M06-2X functionals to ensure robustness and accuracy in describing the reaction energetics and transition states of the crosslinking process.[51–53] The basis set is chosen to be 6-31+G(d,p).[54–56] Normal mode analysis is carried out at the same level of theory to confirm that all optimized geometries correspond to local minima by ensuring positive vibrational frequencies. Transition states (TSs) are verified by the presence of only one imaginary frequency in the normal mode analysis. Intrinsic reaction coordinate calculations are further performed to establish the reaction pathway that connects reactants, products, and TSs following the reaction coordinates. Thermochemical corrections are performed to obtain temperature-corrected Gibbs free energy, $G$ at 413.15 K, which corresponds to the experimentally observed onset of crosslinking.[39] These corrections include zero-point vibrational energy (ZPE), as well as thermal enthalpy and entropy contributions that are derived from the vibrational frequency analysis. The total Gibbs free energy is calculated using the relation[57]:

$$G = E_{elec} + ZPE + \Delta H_{thermal} - T \cdot S_{thermal}$$

Where, $E_{elec}$ is the electronic energy at 0 K, ZPE is the zero-point vibrational energy, and $\Delta H_{thermal}$, is the thermal enthalpy and entropy contributions. The $S_{thermal}$ is the total entropy of the system at the target temperature (T). Thermal corrections to enthalpy and entropy are evaluated from the translational, rotational, and vibrational partition functions using the rigid rotor and harmonic oscillator approximations, assuming ideal gas behaviour. Reaction kinetics are analyzed by calculating the rate constant using the Eyring equation[58]:

$$k = \frac{k_B T}{h} e^{\frac{-\Delta G^{\ddagger}}{RT}}$$

Where k is the rate constant, $k_B$ is the Boltzmann constant, $T$ is the temperature, h is the Planck constant, $\Delta G^{\ddagger}$ is the free energy barrier obtained from Gaussian output, and R is the gas constant. The reaction time is estimated as the inverse of the rate constant (1/$k$), assuming first-order kinetics.

## III. Results and Discussion

All the optimized structures along with critical bond lengths and bond angles are presented in Figure 1 as obtained from the DFT calculations at the M06-2X/6-31+G(d,p) level of theory.

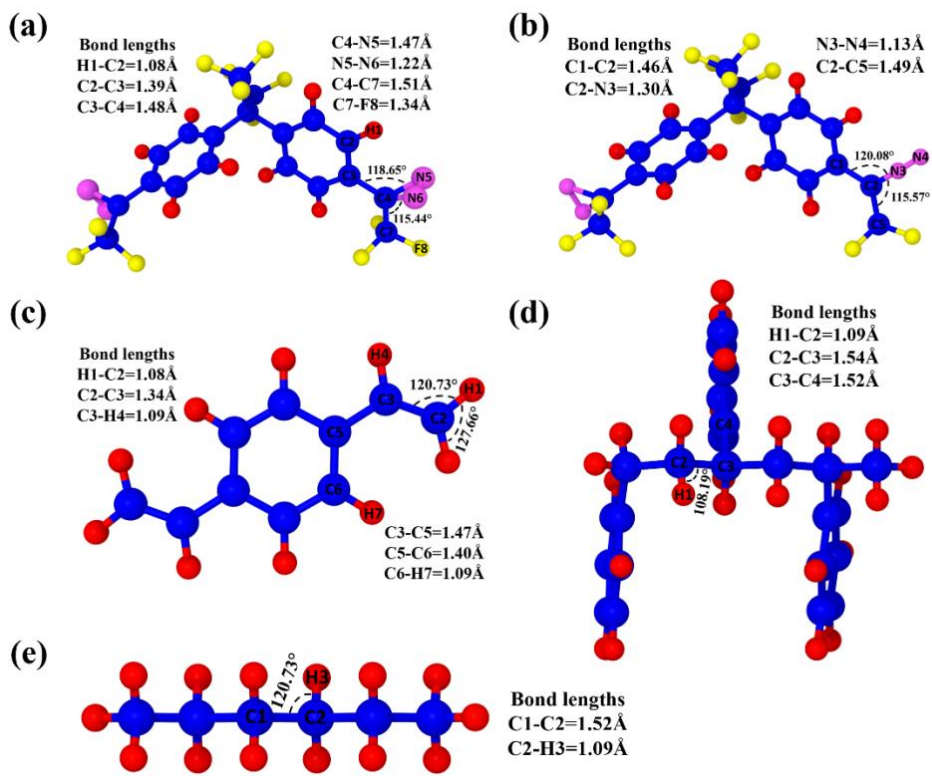

*Figure 1: Optimized structures obtained using M06-2X/6-31G+(d, p) level of theory: Optimized structure of (a) bis-diazirine (BD) crosslinker (b) diazo-isomer (c) divinylbenzene (DVB) (d) polystyrene (PS) three repeat units (e) polyethylene (PE) three repeat units. Blue, red, yellow, and pink spheres represent carbon (C), hydrogen (H), fluorine (F), and nitrogen (N) atoms, respectively.*

For BD (Figure 1a), the C2–C3 bond (1.39 Å) confirms the double bond character, while the longer C4–C7 bond (1.51 Å) reflects electronic effects from the adjacent electronegative fluorine. The N5–N6 bond (1.22 Å) indicates a diazirene (-N=N-) linkage. Notable angles include C3–C4–N5 (118.65°) and C7–C4–N6 (115.44°) around the reaction centre. In the diazo-isomer (Figure 1b), the C2–N3 (1.30 Å) and N3–N4 (1.13 Å) bonds confirm the formation of a diazo group. The bond angles near the reaction centre are N3–C2–C1 (120.08°) and N3–C2–C5 (115.57°). Similarly, Figure 1c shows important bond lengths and bond angles in DVB near the reaction centre. The PS structure (Figure 1d) shows backbone elongation due to bulky phenyl groups: C2–C3 is 1.54 Å, and C3–C4 is 1.52 Å, with a reduced angle of 108.19° at H1–C2–C3. In contrast, PE structure (Figure 1e) displays uniform saturated hydrocarbon geometry: C1–C2 bond is 1.52 Å, C2–H3 is 1.09 Å, and the bond angle C1–C2–H3 is 120.73°, indicating a relatively strain-free backbone. These DFT-optimized structures of reactants and products are utilized to examine the reaction pathways.

Two possible pathways for the crosslink formation are identified. Figure 2a and c correspond the crosslink formation between two PE chains; and Figure 2b and d correspond to the crosslink formation between two PS chains. In both cases, a BD molecule is used as the crosslinker.

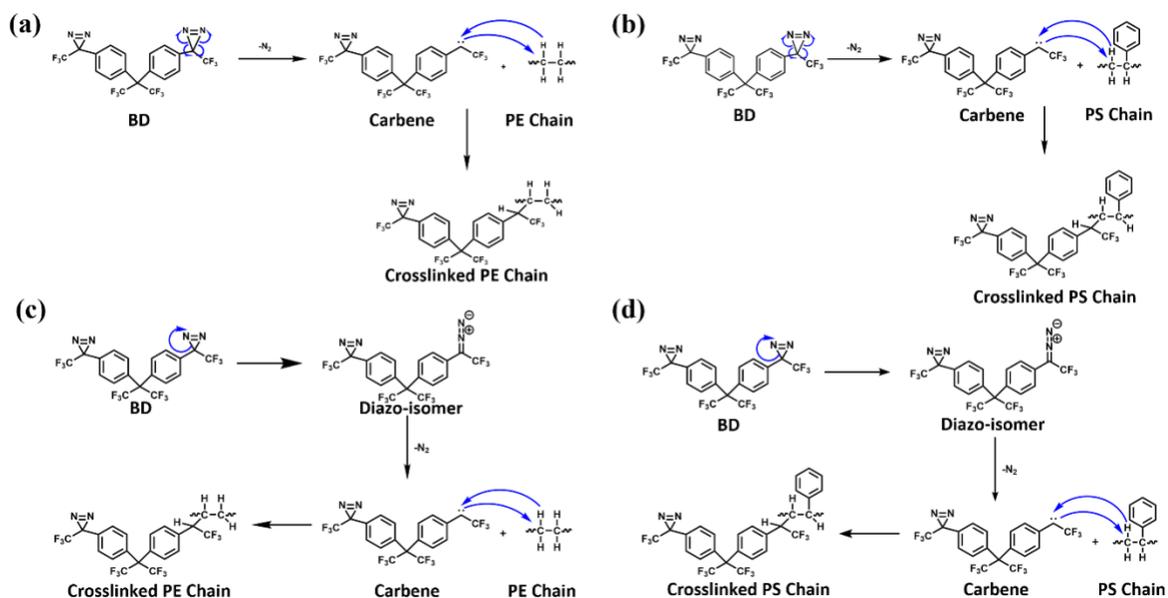

*Figure 2: Reaction mechanisms for the crosslinking of bis-diazirine (BD) crosslinker with polymer chains: Reaction pathway 1 for (a) PE crosslinking, (b) PS crosslinking, and Reaction pathway 2 for (c) PE crosslinking and (d) PS crosslinking.*

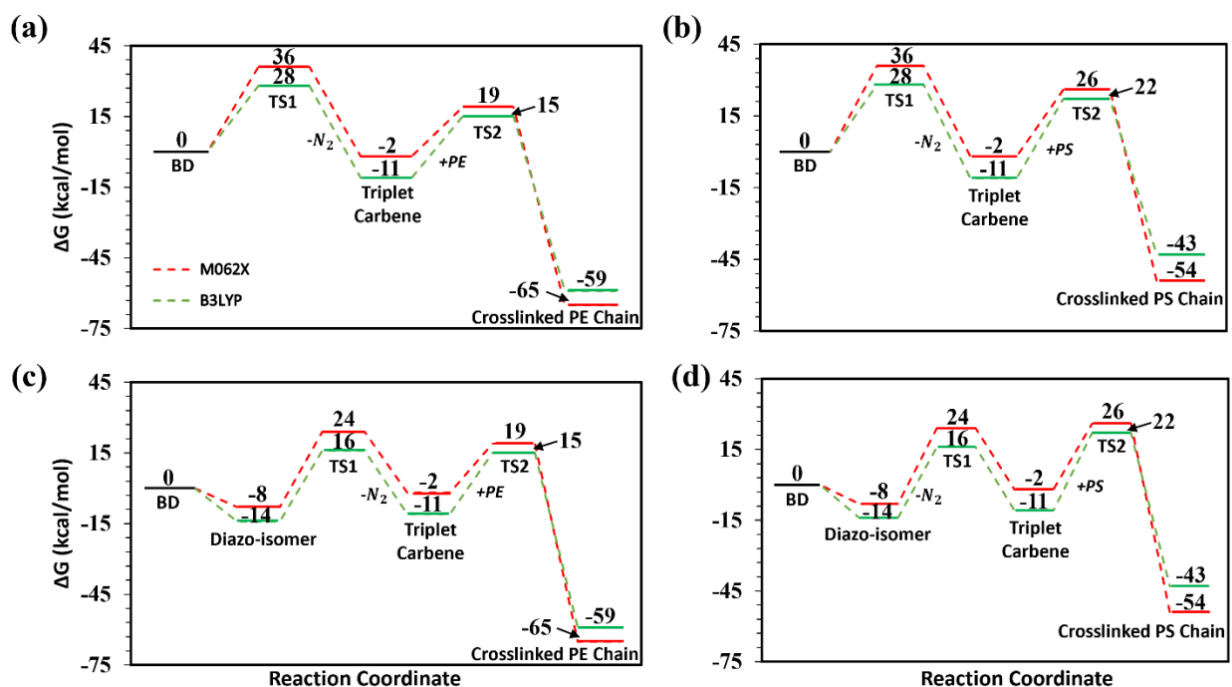

*Figure 3: Relative free energy profiles for the crosslinking reaction of BD with polymer trimers computed using M06-2X (red dashed line) and B3LYP (green dashed line): Reaction pathway 1 for (a) PE crosslinking (b) PS crosslinking and Reaction pathway 2 for (c) PE crosslinking and (d) PS crosslinking.*

The corresponding relative free energy (ΔG) profiles, computed using M06-2X and B3LYP level of theory, are shown in Figure 3. In the reaction pathway-1 (Figures 2a and 3a), the crosslinking process proceeds with the elimination of nitrogen gas ($N_2$) from the BD crosslinker through the first transition state (TS1), and the free energy barrier for this step is 36 kcal/mol at the M06-2X level of theory. This step leads to the formation of a highly reactive carbene intermediate. Possibility of both singlet and triplet states are examined, and our calculations indicate that the triplet carbene is more stable than the singlet carbene, with a relative free energy of –2 kcal/mol with respect to BD. The triplet carbene then abstracts a hydrogen atom from the PE chain, creating a radical site on the polymer. This radical formation facilitates the subsequent C–C bond formation between the BD and the PE chain by lowering the free energy barrier, which corresponds to the crosslink formation and is associated with a free energy barrier (TS2) of 21 kcal/mol. The final crosslinked PE chain is highly stabilized, with a relative free energy of –65 kcal/mol, indicating a strongly exothermic and thermodynamically favourable reaction. Among all steps, the nitrogen elimination step (TS1) has the highest energy barrier, making it the rate-determining step. An alternative reaction, pathway-2 (Figures 2c & 3c), involves the formation of a diazo-isomer intermediate via an electronic rearrangement of the diazo moity in the BD molecule. This intermediate is more stable with respect to BD by –8 kcal/mol. The diazo-isomer then undergoes N$_2$ elimination via TS1, with a free energy barrier of 32 kcal/mol, forming the same triplet carbene as in the pathway-1. The subsequent steps, including hydrogen abstraction and crosslinking via TS2, proceed similarly to those in pathway-1. The presence of a more stable diazo intermediate and a lower TS1 barrier makes the pathway-2 energetically more favourable than the pathway-1. A similar reaction mechanism is observed in the case of PS crosslinking. However, in the reaction pathway-1 for PS crosslinking (Figures 2b and 3b), the second transition state (TS2) corresponding to the C–C bond formation is slightly elevated at 28 kcal/mol, compared to the PE system (21 kcal/mol) due to the presence of the bulky phenyl rings in the PS chain. The final crosslinked PS structure is also stabilized, with a relative free energy of –54 kcal/mol. Reaction pathway-2 for PS crosslinking (Figures 2d and 3d) follows the same mechanism as PE chain with formation of a diazo-isomer intermediate and a slightly increased free energy barrier (28 kcal/mol) for TS2, leading to a final crosslinked product at –54 kcal/mol. We also perform these calculations using the B3LYP functional. Our calculations show that the relative free energy values are slightly lower compared to the M06-2X functional, except for the final crosslinked products, where relative free energies computed using the B3LYP functional are found to be slightly higher (-59 kcal/mol for the PE case and -43 kcal/mol for the PS case). This difference can be attributed to the B3LYP's less rigorous treatment of dispersion and mid-range correlation effects. However, the overall reaction mechanisms and the relative energetics associated with the crosslinking pathways remain consistent across both levels of theory. Our study finds the presence of a carbene triplet state that is lower in energy than the singlet state. This result shows that the carbene intermediate formed after N$_2$ elimination can transition from singlet to triplet. The triplet state is more stable, with a singlet-triplet energy difference of 8 kcal/mol and 7 kcal/mol using M06-2X and B3LYP level of theory, respectively. Difference in energy is significantly higher than thermal energy at room temperature (~0.6 kcal/mol). This result aligns with earlier findings by Musolino et al., who reported that trifluoromethyl aryl diazirines primarily yield triplet carbenes as a ground-state intermediate.[59]

To gain a deeper understanding into the crosslinking mechanism of the BD crosslinker across various polymer systems, we evaluate the free energy barriers for BD crosslinker C–H insertion step (TS2 in Figure 3) at different reactive carbon centres in the polymer chain—namely, secondary (2°) and tertiary (3°) carbons—within polymer systems of PE, PP, and PS. The computed free energy barriers are shown in Figure 4. In conventional radical crosslinking mechanisms of two polymer chains of PP or PS by peroxides or irradiation, tertiary carbon sites are generally more reactive due to the lower bond dissociation energy of tertiary C–H bonds and the enhanced stability of the resulting tertiary radicals.[47,60] However, for molecular crosslinker, the free energy barrier to crosslink tertiary carbon and that for secondary carbon is not well known. Interestingly, our DFT calculations reveal a contrasting trend when employing BD as a molecular crosslinker. For PE, the C-H insertion at the secondary carbon has the lowest free energy barrier of 21 kcal/mol among all the systems at the M06-2X level of theory. This is evident as the linear structure of PE chain makes it easier for the carbene formed in the BD molecule to abstract the hydrogen from the C-H bond in a PE chain and form the crosslink. In PP, the free energy barrier for insertion at the tertiary carbon is slightly lower (34 kcal/mol) compared to the secondary carbon (38 kcal/mol). This is consistent with the balance between favourable radical stabilization at the 3° centre and modest steric crowding from adjacent methyl groups. Conversely, the secondary carbon of PS is more reactive, with a free energy barrier of 28 kcal/mol, while the tertiary carbon shows a significantly higher free energy barrier of 38 kcal/mol. This inversion is attributed to significant steric hindrance from the bulky phenyl group attached to the tertiary carbon, which raises the energy barrier for C-H insertion and reduces reactivity, though tertiary radicals are generally more stable. Calculations of free energy barrier at the B3LYP level show similar trends as that of the M06-2X level theory, as shown in Figure 4.

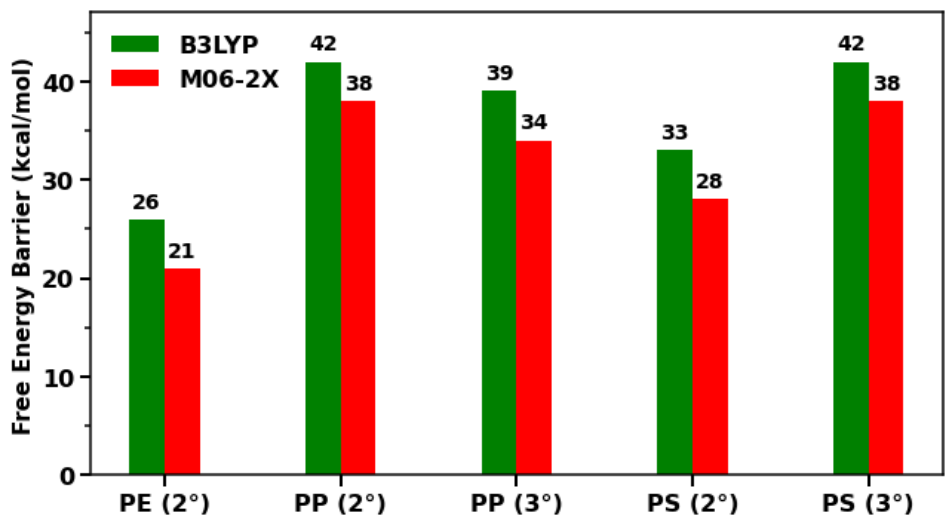

*Figure 4: Free energy barriers for BD crosslinker C-H insertion into secondary (2°) and tertiary (3°) carbons in PE, PP, and PS three repeat unit systems, as computed using the B3LYP (green bar) and M06-2X (red bar) level of theory.*

Next, we investigate the influence of polymer chain length on the C–H insertion step (TS2) during the crosslinking reaction with the BD crosslinker. We carried out a series of calculations using PE chains with varying repeat units. Specifically, we considered chain lengths corresponding to 1-5, 10, and 15 repeat units. All free energy barrier calculations for this analysis were performed at the B3LYP level of theory, given the high computational cost associated with the M06-2X functional. The calculated free energy barrier for the TS2 transition state as a function of polymer chain length is shown in Figure 5. Our calculation shows that overall, the barrier ranges from 25 to 27 kcal/mol, indicating that the influence of polymer chain length on the C–H insertion step is modest (~2 kcal/mol). For shorter chains (1–5 repeat units), the free energy barrier remains relatively constant around 25–26 kcal/mol. A slight increase to 27 kcal/mol is observed at longer chain lengths (10 and 15 repeat units). This subtle increase may be attributed to minor conformational or steric effects that arise with increasing chain length. However, the small magnitude of the change suggests that C–H insertion remains kinetically accessible across the range of chain lengths studied.

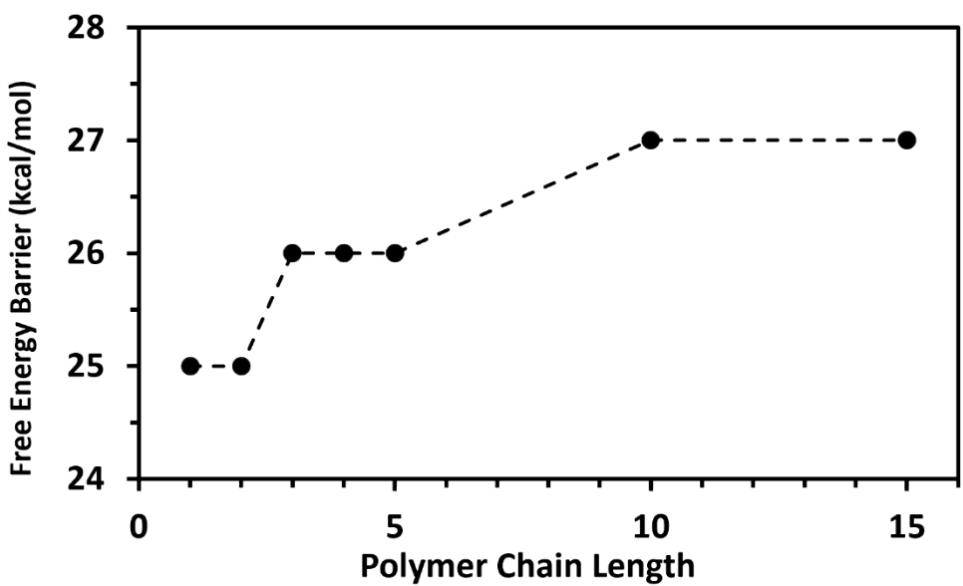

*Figure 5: Free energy barrier as a function of the polymer chain length for the C–H insertion step (TS2) in the crosslinking reaction between the BD and PE. Free energy barrier values are reported for PE chain lengths of 1-5, 10, and 15 repeat units. We use the B3LYP theory for these calculations.*

We now focus the crosslinking reaction mechanism between DVB and polymer chains (PE and PS). The crosslinking reaction mechanism and corresponding relative free energy (ΔG) profiles are shown in Figure 6.

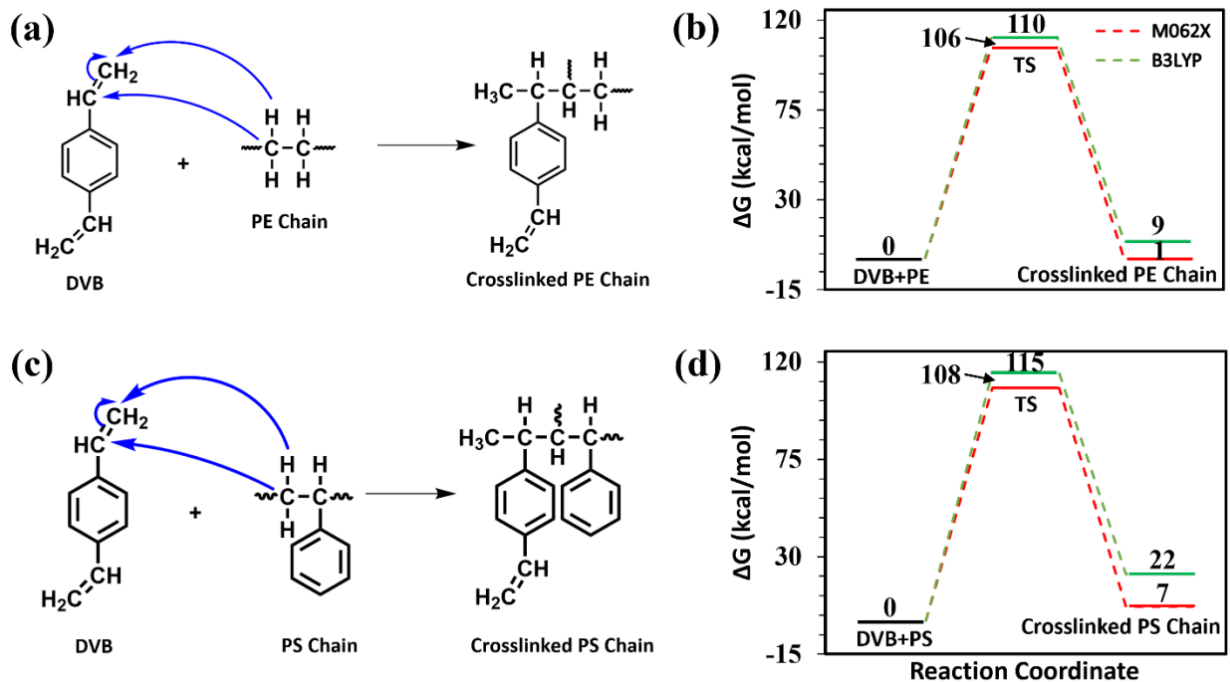

*Figure 6: Crosslink formation mechanism with DVB crosslinker: Reaction mechanism for (a) PE system and (c) PS system. Relative free energy profile for (b) PE system and (d) PS system. Red dashed line represents relative free energy values using M06-2X and green dashed line represents relative free energy values using B3LYP functional.*

The crosslinking reaction between DVB and the polymer chains (PE and PS) follows a radical-mediated mechanism (Figures 6a and 6c). A free radical or thermal activation leads to the opening of the vinyl groups in DVB, generating reactive sites for crosslinking. The activated DVB reacts with the polymer chain by forming a new C-C bond, leading to the crosslinked structure. The reaction stabilizes with the formation of the final crosslinked product. The free energy barrier for the reaction is 106 kcal/mol (M06-2X) for PE case (Figure 6b) which is significantly higher (~5 times) compared to the BD case. The final product is energetically stable, with only a small free energy difference (1 kcal/mol) from the reactant state. The free energy barrier for PS chain crosslinking is slightly higher, 108 kcal/mol for PS case (Figure 6d). This is also significantly higher (~5 times) compared to the BD case. The final product state is energetically higher than the initial reactant state, indicating endothermic reaction with relative free energy value of 7 kcal/mol. The B3LYP level of theory shows consistent higher relative free energy values for the reaction mechanism. These findings suggest a huge energy barrier for the DVB crosslink with polymer, which is impractical at a moderately high temperature. Consistent with this theoretical prediction, recent experimental reports DVB crosslinking with polystyrene at at 343–363 K in the presence of radical initiators such as benzoyl peroxide (BPO).[41] Under such thermal conditions, BPO decomposes to generate free radicals that activate DVB's vinyl groups and facilitate crosslink formation with the polymer matrix.

Table 1 summarizes the calculated free energy barriers and the corresponding reaction times at 413.15 K for the two crosslinking mechanisms involving the molecular crosslinkers BD and DVB. For both systems, the free energy values used in the Eyring equation corresponds to the rate-determining step of the overall crosslinking process. In the case of BD, the rate-determining step (TS1) involves the loss of molecular nitrogen ($N_2$) from the BD crosslinker, leading to the formation of a reactive carbene intermediate. For DVB, the rate-determining step corresponds to the formation of the covalent crosslink between DVB and the polymer chain.

*Table 1: Calculated free energy barriers for rate-determining steps and reaction times for BD and DVB crosslinkers*

| Crosslinker | Polymer | Reaction Pathway | Free Energy Barrier (M062X) (kcal/mol) | Free Energy Barrier (B3LYP) (kcal/mol) | Reaction Time (M062X) | Reaction Time (B3LYP) |
|---|---|---|---|---|---|---|
| **BD** | PE & PS | Pathway 1 | 36 | 28 | 15 Days | 1 Minute |
| **BD** | PE & PS | Pathway 2 | 32 | 30 | 3 Hours | 14 Minutes |
| **DVB** | PE | - | 106 | 110 | $4 \times 10^{35}$ Years | $6 \times 10^{37}$ Years |
| **DVB** | PS | - | 108 | 115 | $5 \times 10^{36}$ Years | $2 \times 10^{40}$ Years |

As discussed earlier, for BD crosslinker, the reaction proceeds through two pathways, with the $N_2$ removal step identified as the rate-determining step. In Pathway-1, the free energy barrier is 36 kcal/mol (M06-2X) and 28 kcal/mol (B3LYP), leading to a reaction time of 15 days (M06-2X) and 1 minute (B3LYP) respectively. In Pathway-2, the free energy barrier is slightly lower at 32 kcal/mol (M06-2X) and 30 kcal/mol (B3LYP), significantly reducing the reaction time to 3 hours (M06-2X) and 14 minutes (B3LYP). These results clearly indicate that BD crosslinking can occur under mild conditions, without requiring high energy input or catalysts. The optimized reaction time suggests that the crosslinking reaction is practical, allowing polymer chains to crosslink efficiently within a reasonable timeframe. In contrast, DVB crosslinking exhibits significantly higher free energy barriers, leading to extremely long reaction times. For PE, the free energy barrier is calculated to be 106 kcal/mol (M06-2X) and 110 kcal/mol (B3LYP), resulting in reaction times of $4\times10^{35}$ years (M06-2X) and $6\times10^{37}$ years (B3LYP) respectively. For PS, the free energy barrier is 108 kcal/mol (M06-2X) and 115 kcal/mol (B3LYP), leading to reaction times of $5\times10^{36}$ years (M06-2X) and $2\times10^{40}$ years (B3LYP) respectively. These values indicate that DVB crosslinking is essentially non-existent under these conditions, as the reaction would take an impractically long time. Without high energy input or a suitable catalyst, DVB crosslinking is not feasible, as the reaction cannot overcome the substantial free energy barrier. The significant difference in free energy barrier between BD and DVB crosslinkers explains why BD successfully crosslinked the polymer under the given conditions. These findings emphasize the crucial role of the diazirine functional group, which contains a three-membered ring with two nitrogen atoms and one carbon atom. We estimate the energy required to break the ring is very low and that makes the polymer crosslinking very efficient.

## Conclusions

As the demand for high-performance, sustainable, and functional materials grows, innovations in crosslinker design will continue to play an important role in the future of polymer-based technologies. This work provides a first-principles understanding of thermally activated polymer crosslinking mediated by BD crosslinkers, with comparative insight into traditional DVB-based crosslink mechanisms. Through comprehensive DFT calculations at both M06-2X and B3LYP levels, we elucidate the energetics and mechanistic pathways of C–H insertion reactions across three polymer matrices viz., PE, PP and PS. Our findings demonstrate that BD crosslinkers offer significantly lower free energy barriers and faster reaction kinetics than DVB—especially on saturated polyolefins where conventional crosslinkers fail to react efficiently. Notably, the carbene intermediate derived from BD exhibits selective reactivity towards secondary C–H bonds in PE and PS due to favorable geometric accessibility and minimized steric hindrance, while in PP, tertiary sites are slightly more favorable due to radical stabilization effects. The computed reaction barriers remain moderately low (~21–38 kcal/mol), enabling activation at modest thermal conditions (383–413 K) without the need for external catalysts. Furthermore, the effect of polymer chain length on C–H insertion is found to be minimal, suggesting robust applicability of BD crosslinkers across a wide range of polymer architectures. Overall, this study establishes BD as a versatile and efficient crosslinking agent capable of engaging unfunctionalized C–H bonds under mild thermal stimuli. These insights not only rationalize the experimental success of BD-based systems but also highlight a paradigm shift from functional group-specific crosslinking to universal C–H activation, opening new avenues in polymer modification and material design.

## Acknowledgements

This work is made possible by financial support from the SERB, DST, Govt of India through a core research grant (CRG/2022/006926).

## References:

1. Jeremic, D. Polyethylene. in *Ullmann's Encyclopedia of Industrial Chemistry* (ed. Wiley-VCH Verlag GmbH & Co. KGaA) 1–42 (Wiley-VCH Verlag GmbH & Co. KGaA, Weinheim, Germany, 2014). doi:10.1002/14356007.a21_487.pub3.

2. Song, P. & Wang, H. High-Performance Polymeric Materials through Hydrogen-Bond Cross-Linking. *Advanced Materials* **32**, 1901244 (2020).

3. Tillet, G., Boutevin, B. & Ameduri, B. Chemical reactions of polymer crosslinking and post-crosslinking at room and medium temperature. *Progress in Polymer Science* **36**, 191–217 (2011).

4. Xu, Z. *et al.* Solvent-Resistant Self-Crosslinked Poly(ether imide). *Macromolecules* **54**, 3405–3412 (2021).

5. Whiteley, K. S., Heggs, T. G., Koch, H., Mawer, R. L. & Immel, W. Polyolefins. in *Ullmann's Encyclopedia of Industrial Chemistry* (John Wiley & Sons, Ltd, 2000). doi:10.1002/14356007.a21_487.

6. Sperling, L. H. *Introduction to Physical Polymer Science*. (Wiley, 2005). doi:10.1002/0471757128.

7. Heath, D. E. & Cooper, S. L. Polymers. in *Biomaterials Science* 64–79 (Elsevier, 2013). doi:10.1016/B978-0-08-087780-8.00008-5.

8. *Physical Properties of Polymers Handbook*. (Springer, New York, NY, 2007). doi:10.1007/978-0-387-69002-5.

9. Grabska-Zielińska, S. Cross-Linking Agents in Three-Component Materials Dedicated to Biomedical Applications: A Review. *Polymers* **16**, 2679 (2024).

10. Abbasi, R., Mitchell, A., Jessop, P. G. & Cunningham, M. F. Crosslinking $CO_2$-switchable polymers for paints and coatings applications. *RSC Appl. Polym.* **2**, 214–223 (2024).

11. Di, L. *et al*. Influence of Crosslink Density on Electrical Performance and Rheological Properties of Crosslinked Polyethylene. *Polymers* **16**, 676 (2024).

12. Fan, W., Du, Y., Yuan, Z., Zhang, P. & Fu, W. Cross-Linking Behavior and Effect on Dielectric Characteristics of Benzocyclobutene-Based Polycarbosiloxanes. *Macromolecules* **56**, 6482–6491 (2023).

13. Guan, T. *et al*. Mechanically Robust Skin-like Poly(urethane-urea) Elastomers Cross-Linked with Hydrogen-Bond Arrays and Their Application as High-Performance Ultrastretchable Conductors. *Macromolecules* **55**, 5816–5825 (2022).

14.     He, D. *et al.* Research on Mechanical, Physicochemical and Electrical Properties of XLPE-Insulated Cables under Electrical-Thermal Aging. *Journal of Nanomaterials* **2020**, 1–13 (2020).

15.     Akiba, M. Vulcanization and crosslinking in elastomers. *Progress in Polymer Science* **22**, 475–521 (1997).

16.     Definitions of terms relating to the structure and processing of sols, gels, networks, and inorganic-organic hybrid materials (IUPAC Recommendations 2007). *Pure and Applied Chemistry* (2007).

17.     Bansal, N. & Arora, S. Exploring the impact of gamma rays and electron beam irradiation on physico-mechanical properties of polymers & polymer composites: A comprehensive review. *Nuclear Instruments and Methods in Physics Research Section B: Beam Interactions with Materials and Atoms* **549**, 165297 (2024).

18.     Melilli, G., Guigo, N., Robert, T. & Sbirrazzuoli, N. Radical Oxidation of Itaconic Acid-Derived Unsaturated Polyesters under Thermal Curing Conditions. *Macromolecules* **55**, 9011–9021 (2022).

19.     Ravve, A. *Principles of Polymer Chemistry*. (Springer, New York, NY, 2012). doi:10.1007/978-1-4614-2212-9.

20.     Melo, R. P. D., Aguiar, V. D. O. & Marques, M. D. F. V. Silane Crosslinked Polyethylene from Different Commercial PE's: Influence of Comonomer, Catalyst Type and Evaluation of HLPB as Crosslinking Coagent. *Mat. Res.* **18**, 313–319 (2015).

21. Celina, M. C. Review of polymer oxidation and its relationship with materials performance and lifetime prediction. *Polymer Degradation and Stability* **98**, 2419–2429 (2013).

22. Hamid, S. H. *Handbook of Polymer Degradation*. (CRC Press, Boca Raton, 2000). doi:10.1201/9781482270181.

23. Bhattacharya, A. Radiation and industrial polymers. *Progress in Polymer Science* **25**, 371–401 (2000).

24. Chmielewski, A. G., Haji-Saeid, M. & Ahmed, S. Progress in radiation processing of polymers. *Nuclear Instruments and Methods in Physics Research Section B: Beam Interactions with Materials and Atoms* **236**, 44–54 (2005).

25. *Handbook of Polyolefins*. (CRC Press, Boca Raton, 2000). doi:10.1201/9780203908716.

26. Braun, D., Cherdron, H., Rehahn, M., Ritter, H. & Voit, B. *Polymer Synthesis: Theory and Practice: Fundamentals, Methods, Experiments*. (Springer Berlin Heidelberg, Berlin, Heidelberg, 2013). doi:10.1007/978-3-642-28980-4.

27. Kroll, D. M. & Croll, S. G. Influence of crosslinking functionality, temperature and conversion on heterogeneities in polymer networks. *Polymer* **79**, 82–90 (2015).

28. Tillet, G., Boutevin, B. & Ameduri, B. Chemical reactions of polymer crosslinking and post-crosslinking at room and medium temperature. *Progress in Polymer Science* **36**, 191–217 (2011).

29. Shundo, A., Aoki, M., Yamamoto, S. & Tanaka, K. Cross-Linking Effect on Segmental Dynamics of Well-Defined Epoxy Resins. *Macromolecules* **54**, 5950–5956 (2021).

30. Shundo, A., Yamamoto, S. & Tanaka, K. Network Formation and Physical Properties of Epoxy Resins for Future Practical Applications. *JACS Au* **2**, 1522–1542 (2022).

31. Kunnikuruvan, S., Parandekar, P. V., Prakash, O., Tsotsis, T. K. & Nair, N. N. Polymerization Mechanism and Cross-Link Structure of Nadic End-Capped Polymers: A Quantum Mechanical and Microkinetic Investigation. *Macromolecules* **50**, 6081–6087 (2017).

32. Lattuada, M. *et al.* Kinetics of Free-Radical Cross-Linking Polymerization: Comparative Experimental and Numerical Study. *Macromolecules* **46**, 5831–5841 (2013).

33. Wang, Z. J. & Gong, J. P. Mechanochemistry for On-Demand Polymer Network Materials. *Macromolecules* **58**, 4–17 (2025).

34. Brunner, J., Senn, H. & Richards, F. M. 3-Trifluoromethyl-3-phenyldiazirine. A new carbene generating group for photolabeling reagents. *Journal of Biological Chemistry* **255**, 3313–3318 (1980).

35. Blencowe, A., Blencowe, C., Cosstick, K. & Hayes, W. A carbene insertion approach to functionalised poly(ethylene oxide)-based gels. *Reactive and Functional Polymers* **68**, 868–875 (2008).

36. Welle, A., Billard, F. & Marchand-Brynaert, J. Tri- and Tetravalent Photoactivable Cross-Linking Agents. *Synthesis* **44**, 2249–2254 (2012).

37. BURGOON, H., CYRUS, C. D. & Rhodes, L. F. Diazirine compounds as photocrosslinkers and photoimageable compositions comprising them. (2016).

38. Rhodes, L. F., Burgoon, H., Afonina, I., Backlund, T. & Morley, A. Diazirine containing organic electronic compositions and device thereof. (2019).

39. Lepage, M. L. *et al.* A broadly applicable cross-linker for aliphatic polymers containing C–H bonds. *Science* **366**, 875–878 (2019).

40. Okay, O., Kurz, M., Lutz, K. & Funke, W. Cyclization and Reduced Pendant Vinyl Group Reactivity during the Free-Radical Crosslinking Polymerization of 1,4-Divinylbenzene. *Macromolecules* **28**, 2728–2737 (1995).

41. Sajjadi, S., Keshavarz, S. A. M. & Nekoomanesh, M. Kinetic investigation of the free-radical crosslinking copolymerization of styrene with a mixture of divinylbenzene isomers acting as the crosslinker. *Polymer* **37**, 4141–4148 (1996).

42. Dong, J. Y., Hong, H., Chung, T. C., Wang, H. C. & Datta, S. Synthesis of Linear Polyolefin Elastomers Containing Divinylbenzene Units and Applications in Cross-Linking, Functionalization, and Graft Reactions. *Macromolecules* **36**, 6000–6009 (2003).

43. Dong, J. Y., Hong, H., Chung, T. C., Wang, H. C. & Datta, S. Synthesis of Linear Polyolefin Elastomers Containing Divinylbenzene Units and Applications in Cross-Linking, Functionalization, and Graft Reactions. *Macromolecules* **36**, 6000–6009 (2003).

44. Koh, M. L., Konkolewicz, D. & Perrier, S. A Simple Route to Functional Highly Branched Structures: RAFT Homopolymerization of Divinylbenzene. *Macromolecules* **44**, 2715–2724 (2011).

45. Dale, J. A. & Millar, J. R. Crosslinker effectiveness in styrene copolymerization. *Macromolecules* **14**, 1515–1518 (1981).

46. Okay, O. Macroporous copolymer networks. *Progress in Polymer Science* **25**, 711–779 (2000).

47. Rätzsch, M., Arnold, M., Borsig, E., Bucka, H. & Reichelt, N. Radical reactions on polypropylene in the solid state. *Progress in Polymer Science* **27**, 1195–1282 (2002).

48. Wilson, M. E., Wilson, J. A. & Kurth, M. J. Enhanced Site Isolation in Cross-Link-Functionalized Polystyrene Networks:  Mobility Studies Using Steady-State Fluorescence and ESR Techniques. *Macromolecules* **30**, 3340–3348 (1997).

49. Dennington, R., Keith, T. A. & Millam, J. M. GaussView, version 6.0. 16. *Semichem Inc Shawnee Mission KS* **13**, (2016).

50. Frisch, M. J. *et al.* G16_C01. Gaussian 16, Revision C. 01, Gaussian. *Inc., Wallin* **248**, (2016).

51. Becke, A. D. Density-functional thermochemistry. III. The role of exact exchange. *The Journal of Chemical Physics* **98**, 5648–5652 (1993).

52. Lee, C., Yang, W. & Parr, R. G. Development of the Colle-Salvetti correlation-energy formula into a functional of the electron density. *Phys. Rev. B* **37**, 785–789 (1988).

53. Zhao, Y. & Truhlar, D. G. The M06 suite of density functionals for main group thermochemistry, thermochemical kinetics, noncovalent interactions, excited states, and transition elements: two new functionals and systematic testing of four M06-class functionals and 12 other functionals. *Theor Chem Account* **120**, 215–241 (2008).

54. Hehre, W. J., Ditchfield, R. & Pople, J. A. Self—Consistent Molecular Orbital Methods. XII. Further Extensions of Gaussian—Type Basis Sets for Use in Molecular

Orbital Studies of Organic Molecules. *The Journal of Chemical Physics* **56**, 2257–2261 (1972).

55. Hariharan, P. C. & Pople, J. A. The influence of polarization functions on molecular orbital hydrogenation energies. *Theoret. Chim. Acta* **28**, 213–222 (1973).

56. Clark, T., Chandrasekhar, J., Spitznagel, G. W. & Schleyer, P. V. R. Efficient diffuse function-augmented basis sets for anion calculations. III. The 3-21+G basis set for first-row elements, Li–F. *J Comput Chem* **4**, 294–301 (1983).

57. McQuarrie, D. A., Simon, J. D. & Simon, J. D. *Molecular Thermodynamics*. (University Science Books, Sausalito, Calif, 1999).

58. Eyring, H. The Activated Complex in Chemical Reactions. *The Journal of Chemical Physics* **3**, 107–115 (1935).

59. Musolino, S. F., Pei, Z., Bi, L., DiLabio, G. A. & Wulff, J. E. Structure–function relationships in aryl diazirines reveal optimal design features to maximize C–H insertion. *Chem. Sci.* **12**, 12138–12148 (2021).

60. Guyot, A. Recent Developments in the Thermal Degradation of Polystyrene-A Review. *Polymer Degradation and Stability* **15**, 219–235 (1986).